\documentclass[aip,pof,preprint,amsmath,amssymb]{revtex4-1} 
\usepackage{graphicx}  
\usepackage{dcolumn}   
\usepackage{bm}        
\usepackage{amssymb}   
\usepackage{color}
\usepackage{tabularx}
\usepackage{booktabs}
\usepackage{amsmath}
\usepackage{afterpage}
\usepackage{color}

\begin{document}
\title{Spectra and probability distributions of  thermal flux in turbulent Rayleigh-B\'{e}nard convection}
\author{Hirdesh K. Pharasi}
\affiliation{Department of Physics, Doon University, Dehradun-248 001, India }
\author{Deepesh Kumar}
\affiliation{Department of Chemical and Materials Engineering, University of Alberta, Edmonton AB-T6G 2V4, Canada}
\author{Krishna Kumar}
\affiliation{Department of Physics, Indian Institute of Technology, Kharagpur-721 302, India}
\author{Jayanta K. Bhattacharjee}
\affiliation{Harish-Chandra Research Institute, Allahabad-211 019, India} 
\date{\today}
\begin{abstract}
The spectra of turbulent heat flux $\mathrm{H}(k)$ in Rayleigh-B\'{e}nard convection with and without uniform rotation are presented. The spectrum $\mathrm{H}(k)$ scales with wave number $k$ as $\sim k^{-2}$. The scaling exponent is almost independent of the Taylor number $\mathrm{Ta}$ and Prandtl number $\mathrm{Pr}$  for higher values of the reduced Rayleigh number $r$ ($> 10^3$).  The exponent, however, depends on $\mathrm{Ta}$ and $\mathrm{Pr}$ for smaller values of $r$ ($<10^3$). The probability distribution functions of the local heat fluxes are non-Gaussian and have exponential tails.
\end{abstract}
\pacs{47.27.te, 47.27.ek, 47.27.er}
\maketitle
\section{Introduction}
The study of fully developed turbulence in  stratified fluids produced by an
externally maintained temperature gradient along the direction of gravitational field has generally involved two different kinds of investigations. One has been the study of the Nusselt number\cite{review_RBC1, cioni_etal_1997, castaing_1989, niemela_etal_2000, grossmann_lohse_2000}, which is a measure of heat flux- the rate of transfer of heat from the bottom plate to the top plate  when the fluid is enclosed in a box heated from below. The primary transport is due to turbulent convection and hence the Nusselt number ($\mathrm{Nu}$) is a probe of turbulence in fluids. The other subject of study has been the wave-number dependent energy and entropy spectra~\cite{K41, BO, review_RBC2, ashkenazi_steinberg_1999, shang_xia_2001, zhou_xia_2001, calzavarini_etal_2002,
kunnen_etal_pre_2008, mishra_etal_pre_2010}, which give a picture of how the kinetic energy per unit mass ($E = \frac{1}{2V} \int{ v^2  d^3r}$) and the entropy ($S = \frac{1}{2V} \int{ (\delta T)^2 d^3r}$) are distributed in the wave-number space for a fluid enclosed in a volume $V$. The spectra $E (k)$ and $S(k)$ are defined as $E = \int{E(k) dk}$ and $S = \int{S(k) dk}$, where $k$ is the wave number. The extensive experimental and numerical works have dealt with the Rayleigh number ($\mathrm{Ra}$) dependence of the Nusselt number~\cite{review_RBC1, cioni_etal_1997, castaing_1989, niemela_etal_2000, grossmann_lohse_2000} and the scaling behavior of $E (k)$ and $S (k)$ in the wave-number space~\cite{K41, BO, review_RBC2, ashkenazi_steinberg_1999, shang_xia_2001, zhou_xia_2001, calzavarini_etal_2002, kunnen_etal_pre_2008, mishra_etal_pre_2010}. In the first part of this work,  a bridge is constructed between the two approaches  by studying the spectrum of the heat flux. In analogy with $E (k)$ and $S (k)$, the spectrum $\mathrm{H}(k)$ of the heat flux is defined by the relation $\mathrm{Nu}-1 = \int{\mathrm{H}(k) dk}$. Two decades ago, the heat flux spectrum $\mathrm{H}(k)$ had once been studied by Kerr~\cite{kerr_1996} but no systematic  data showing the scaling behavior in the wave-number space was obtained. Simulations show that $\mathrm{H}(k) \propto k^{-2}$ and the change in the scaling exponent is also studied when the system is subjected to uniform rotation. The data with uniform rotation are confined to Rossby numbers ($\mathrm{Ro}$) greater than unity. In the second part of this work, a study of the probability distribution functions (PDFs) for the heat flux is presented. This study is specifically important because the probability distribution gives information about all the moments and further the PDF for the vertical heat flux should be asymmetric about the origin. It should also have a tail varying much slower than a Gaussian tail for positive (upward) flux, showing the importance of rare events. The PDFs for the heat fluxes in the horizontal plane, on the other hand, should be symmetric about the origin. The PDFs of heat fluxes at finite rotation speeds at different Prandtl numbers have been investigated. Exponential tails are observed for PDFs of heat fluxes. The heat flux in the vertical direction is found to be strongly asymmetrical about the maximum located at the origin, while heat fluxes in the horizontal plane are found to be symmetric. 

In the study of the energy and entropy spectra there has been a puzzle for the last two decades. The homogeneous isotropic turbulence is governed by the Kolmogorov scaling~\cite{K41}. However, there are two competing scenarios in stratified fluids - The Kolmogorov variety and another due to Bolgiano~\cite{BO} and Obukhov~\cite{review_RBC2}. In the Kolmogorov scenario both $E (k)$ and $S (k)$ are supposed to scale as $k^{-5/3}$, whereas in Bolgiano-Obukhov case one has $E (k)\propto k^{-11/5}$ and $S(k) \propto k^{-7/5}$. Interestingly enough there has never been  any unambiguous determination of $E (k)$ and as for $S(k)$ the results have always favored Bolgiano scaling. A clear cut crossover has only been observed only in frequency space~\cite{pharasi_etal_pre2_2014}. The issue of scale invariance of the governing equations has been also discussed in general terms and on the basis of that it is concluded that the scaling for the Bolgiano-Obukhov scenario is: $\mathrm{H}(k) \propto k^{-9/5}$. It is argued that the observed spectrum is actually consistent with the Bolgiano view point.

The results presented in Sec. II are for a situation where scaling is expected to hold in the inertial range. The effects of uniform rotation are investigated in Sec. III. The rotation introduces the Coriolis force in the horizontal plane and in addition to introducing an additional anisotropy, it also breaks the scale invariance of the system. The known results for the rotating convection primarily concern the Nusselt number and one of the important observation was the deviation from the approximate $\mathrm{Ra}^{2/7}$ scaling~\cite{ashkenazi_steinberg_1999, shang_xia_2001, zhou_xia_2001, calzavarini_etal_2002, kunnen_etal_pre_2008, mishra_etal_pre_2010}  as the Rayleigh number falls below a critical value, which depends strongly on the Taylor number $\mathrm{Ta}$ and Prandtl number $\mathrm{Pr}$. Below this critical value of $\mathrm{Ra}$, the scaling is $\Omega$-dependent~\cite{king_etal_nature_2009, schmitz_tilgner_2010, pharasi_etal_pre_2011}. The scaling exponent for the thermal flux is found to change with $\Omega$ for lower values of $\mathrm{Ra}$ and it is found to be $\Omega$
independent for large $\mathrm{Ra}$. The large values of $\mathrm{Ra}$ correspond to the range where the Nusselt number $\mathrm{Nu}$ scales as $\mathrm{Ra}^{2/7}$ and is independent of the rotation speed{ (see Fig.~\ref{Nu_Ra}).

\section{Spectra and PDFs of heat fluxes} 
The hydrodynamic system considered here consists of a thin layer of low-Prandtl-number Boussinesq fluid confined between two horizontal plates separated by a distance $d$, which is subjected to a uniform rotation about a vertical axis with angular velocity $\Omega$. Symbols $\alpha$, $\nu$, $\lambda$ and $g$ stand for the thermal expansion coefficient, kinematic viscosity, thermal diffusivity  and acceleration due to gravity, respectively. An  adverse temperature gradient $\beta = (T_1-T_2)/d = \Delta T/d$ is imposed externally in the vertical direction, where $T_1$ and $T_2$ are the temperatures of the bottom and top plates. The convective motion of the Boussinesq fluid is governed by the following dimensionless equations:
\begin{eqnarray}
\partial_t \mathbf{v} + (\mathbf{v}\cdot\boldsymbol{\nabla})\mathbf{v} &=& -\boldsymbol{\nabla}p + \mathrm{Pr}\theta {\mathbf{e}}_3 + {\sqrt{\mathrm{Pr}/\mathrm{Ra}}}\nabla^{2}\mathbf{v} -{\sqrt{\mathrm{Ta} \mathrm{Pr}/ \mathrm{Ra}}} ({\mathbf{e}}_3\mathbf{\times}\mathbf{v}),\label{ns}\\
\mathrm{Pr}\left(\partial_t \theta \mathbf{v}\cdot\boldsymbol{\nabla}\theta\right)  &=& {\sqrt{\mathrm{Pr}/\mathrm{Ra}}} \nabla^{2}\theta + v_3, \label{temp}\\
\boldsymbol{\nabla}\cdot\mathbf{v} &=& 0,\label{cont}
\end{eqnarray}
where $\mathbf{v} (x,y,z,t)\equiv (\mathrm{v}_1, \mathrm{v}_2, \mathrm{v}_3)$,  $p(x,y,z,t)$, and $\theta (x,y,z,t)$ are the deviations
of velocity, pressure and temperature fields, respectively, from their values in the stationary state of conduction.  The unit vector in the vertical upward direction is denoted by ${\bf{e}}_3$. The length, time and temperature is measured in units of fluid thickness $d$, free-fall time $\sqrt{d/(\alpha g \Delta Td)}$, and  characteristic temperature $\nu\Delta T/\lambda$. Rayleigh number $\mathrm{Ra} = g \alpha \beta d^{4}/\lambda \nu$, Prandtl number $\mathrm{Pr} = \nu/\lambda$ and Taylor number $\mathrm{Ta} = 4 \Omega^{2} d^{4}/\nu^{2}$ are three dimensionless numbers characterizing the convective system. Another dimensionless parameter called Rossby number $\mathrm{Ro} = \sqrt{\mathrm{Ra}/(\mathrm{Pr} \mathrm{Ta})}$, which is also used to characterize rotating RBC. The case without rotation corresponds to $\mathrm{Ta} = 0$ (i.e., $\mathrm{Ro} = \infty$).  Boundary conditions for thermally conducting and {\it free-slip} bounding surfaces, located at $z=0$ and $z=1$, are:
\begin{equation}
 \partial_{z}\mathrm{v}_{1} = \partial_{z}\mathrm{v}_{2} = \mathrm{v}_{3}=\theta=0.\label{bcs}
\end{equation}
All the fields are considered periodic in horizontal plane. The expansion of the fields consistent with the boundary conditions [Eq. (\ref{bcs})] are then given by,
\begin{eqnarray}
\mathbf{v}_1 (x,y,z,t) &=& \sum_{l,m,n} U_{lmn}(t) e^{i(lk_xx+mk_yy)} \cos{n\pi z},\label{eq.v1}\\
\mathbf{v}_2 (x,y,z,t) &=& \sum_{l,m,n} V_{lmn}(t) e^{i(lk_xx+mk_yy)} \cos{n\pi z},\label{eq.v2}\\
\mathbf{v}_3 (x,y,z,t) &=& \sum_{l,m,n} W_{lmn}(t) e^{i(lk_xx+mk_yy)} \sin{n\pi z}, \label{eq.v3}\\
\theta(x,y,z,t) &=& \sum_{l,m,n} \Theta_{lmn}(t) e^{i(lk_xx+mk_yy)} \sin{n\pi z},
\label{eq.temp}\\
p(x,y,z,t) &=& \sum_{l,m,n} P_{lmn}(t) e^{i(lk_xx+mk_yy)} \cos{n\pi z}.\label{eq.p}
\end{eqnarray}
The expansion of the pressure field is such that its dependence on the $z$-coordinate is similar to that of the horizontal velocities $\mathrm{v}_1$ and $\mathrm{v}_2$, and its vertical derivative has $z-$ dependence identical to that the vertical velocity $\mathrm{v}_3$.  The pressure modes due to convection is computed by taking divergence of the momentum equation~(\ref{ns}) and using the equation of continuity~(\ref{cont}). 

The hydrodynamic system [Eqs.~(\ref{ns})-(\ref{cont})] with the boundary conditions (Eq.~\ref{bcs}) is numerically integrated using pseudo spectral method. The fourth order Runge-Kutta (RK4) scheme is used for the time advancement. The time steps have been monitored to have Courant-Friedrichs-Lewy (CFL) condition satisfied all the time. The grid resolutions are such that the smallest dissipative (Kolmogorov) scale is well resolved~(see for details~\cite{pharasi_etal_pre_2011}). The grid size $l$ fixes the a cut-off wave number  $k_{max}\sim 1/l$. The viscous dissipative (Kolmogorov) scale is defined as $\eta = (\nu^3 / \epsilon^{max})^{1/4}$, where $\epsilon^{max}$ is the maximum value of the kinetic energy flux.  Similarly, there is another dissipative scale: thermal dissipative scale. It is defined as $\eta_S = (\lambda^3 / \epsilon^{max}_S)^{1/4}$, where $\epsilon^{max}_S$ is the maximum value of the entropy flux. The simulations are monitored to have the minimum values of the products $\eta \times k_{max}$ and  $\eta_S \times k_{max}$ always greater than unity. The average values of this product for the least Kolomogorov scale are listed in Table~\ref{table1}. 

The convection is oscillatory at the onset~\cite{chandrasekhar:book_1961} for $\mathrm{Pr} < 0.67$, if $\mathrm{Ta}$ is greater than a critical value $\mathrm{Ta}_c$.  The reduced Rayleigh number $r$ is therefore defined as $r = \mathrm{Ra}/\mathrm{Ra}_{\circ}$, where  $\mathrm{Ra}_{\circ} (\mathrm{Ta}, \mathrm{Pr})$ is the threshold for oscillatory convection. In the absence of rotation, $r$ is defined as $r = \mathrm{Ra}/\mathrm{Ra}_c$, where $\mathrm{Ra}_c = 27\pi^4/4$. Long signals for more than 150 dimensional time units are computed on $256^3$ spatial grids to determine the probability density functions (PDFs). Using the last values of fields obtained from  the simulation on $256^3$ spatial grids, the final runs are performed on $512^3$ grids points for the computation of the spectra of global thermal flux in the wave-number space. 

As the integration is done in the wave number space, we get at each time step the values of all fields in $k-$ space. The wave number space is divided into several spherical shells. The symbols $\mathrm{H}(k_i)$ represents the convective heat flux in the $i^{th}$ spherical shell of inner radius $k_i$ and outer radius $k_{i+1}$, where $i$ is an integer. The convective heat flux spectrum $\mathrm{H}(k_i)$ are defined as:
\begin{equation}
\mathrm{H}(k_i) = \sum_{k_i {\leq k} < k_{i+1}} [\mathrm{v}_3({\bf k})\theta^*({\bf k}) + \mathrm{v}_3^*({\bf k})\theta({\bf k})].\label{eq.Hk}
\end{equation}
The magnitude of wave vector is 
$k=\left[ k_{\circ}^2 (l^2 + m^2) + \pi^2 n^2\right]^{1/2}$, where 
$k_{\circ} (\mathrm{Pr}, \mathrm{Ta})$ is the critical wave number at the onset of oscillatory convection~\cite{chandrasekhar:book_1961}. All the simulations are done in a box of size $L_x \times L_y\times 1$, where 
$Lx$ $=$ $Ly$ $=$ $2\pi/k_\circ (\mathrm{Ta},\mathrm{Pr})$ for our purposes. This gives us the spectrum $\mathrm{H}(k_i)$ for one instant of time. The spectrum is computed  for several instants of time at equal intervals  over a long period. Then the average of the field data points in the $i^{th}$ spherical shell is evaluated. The sum using Eq.~(\ref{eq.Hk}) gives the time averaged spectrum of the heat flux.  
\begin{table*}[hbt!]
\centering
\caption{\label{table1} List of the Rossby number $\mathrm{Ro} = \sqrt{\mathrm{Ra}/(\mathrm{Pr} \mathrm{Ta})}$, the Prandtl number $\mathrm{Pr}$, the Taylor number $\mathrm{Ta}$, the critical wave number $k_o$ at the onset of oscillatory convection, the critical Rayleigh number $\mathrm{Ra_o}$, the reduced Rayleigh number $r =\mathrm{Ra}/\mathrm{Ra}_{\circ}$, the time averaged value of $\eta \times k_{max}$, and the the range of dimensionless wave numbers for calculating scaling exponent $\delta$. The reduced Rayleigh number for the non-rotating case ($\mathrm{Ta} = 0$) is defined as: $r = \mathrm{Ra}/\mathrm{Ra}_c$ with $\mathrm{Ra}_c = 27 \pi^4/4$.}
\begin{ruledtabular}
\begin{tabular}{ccccccccc}
$\mathrm{Ro}$&$\mathrm{Pr}$&$\mathrm{Ta}$&$k_o(\mathrm{Ta}) $&$\mathrm{Ra_o}(\mathrm{Ta})$ &$r$&$\overline{\eta \times k_{max}}$& Range of $k$ for & Exponent $\delta$ \\ 
& & & & & & &the exponent $\delta$\\\hline
$3.27$& $0.1$& $1.0\times10^6$&$5.50$ & $1.06\times 10^4$& $100$&$8.90$&$9-35$ &$2.52\pm0.19$\\
$3.70$& $0.5$& $1.0\times 10^6$& $8.79$& $6.83\times10^4$ &$100$& $12.05$& $9-60$ &$2.57\pm0.12$ \\
$7.70$& $0.5$& $3.0\times 10^4$& $4.64$ & $8.90\times 10^3$ &$100$&$15.97$& $7-43$ &$2.45\pm0.14$\\
$8.64$& $0.1$& $3.0\times 10^4$& $2.92$& $2.24\times 10^3$ &$100$&$7.60$& $5-35$ &$2.41\pm0.11$\\
$10.26$& $0.5$& $1.0\times 10^4$& $3.78$& $5.26\times 10^3$ &$100$&$15.25$& $9-37$ &$2.38\pm0.17$ \\
$13.27$& $0.1$& $1.0\times 10^4$& $2.54$& $1.76\times 10^3$ &$100$& $7.67$& $5-27$ &$2.35\pm0.18$\\
$20.25$& $0.5$& $1.0\times 10^6$& $8.79$& $6.83\times 10^4$ &$3000$&$3.94$& $10-120$ &$1.98\pm0.05$\\
$23.13$& $0.1$& $1.0\times 10^6$& $5.50$& $1.06\times 10^4$ &$5000$&$2.70$& $8-80$ &$2.07\pm0.07$ \\
$42.19$& $0.5$& $3.0\times 10^4$& $4.64$& $8.90\times 10^3$ &$3000$&$5.45$& $7-76$       &$1.97\pm0.05$ \\
$47.33$& $0.1$& $3.0\times 10^4$& $2.92$& $2.24\times 10^3$ &$3000$&$2.60$& $5-70$ &$1.96\pm0.05$\\
$56.21$& $0.5$& $1.0\times10^4$& $3.78$& $5.26\times 10^3$ &$3000$&$5.20$&  $8-75$ &$1.96\pm0.06$\\
$72.66$& $0.1$& $1.0\times10^4$& $2.54$& $1.76\times 10^3$ &$3000$&$2.65$& $5-51$ &$1.95\pm0.08$\\
$\infty$& $0.1$& $0$&$--$&$--$&$152$ &$7.71$& $4-24$ &$2.22\pm0.22$\\
$\infty$& $0.1$& $0$&$--$&$--$&$3000$&$3.02$& $7-48$ &$1.93\pm0.09$\\
$\infty$& $0.5$& $0$&$--$&$--$&$150$&$15.03$& $5-18$ &$2.29\pm0.31$\\
$\infty$& $0.5$& $0$&$--$&$--$&$3000$&$6.07$& $7-40$ &$1.94\pm0.09$\\
\end{tabular}
\end{ruledtabular}
\end{table*}

\begin{figure*}[htbp!]
\begin{center}
\includegraphics[width=0.9\textwidth]{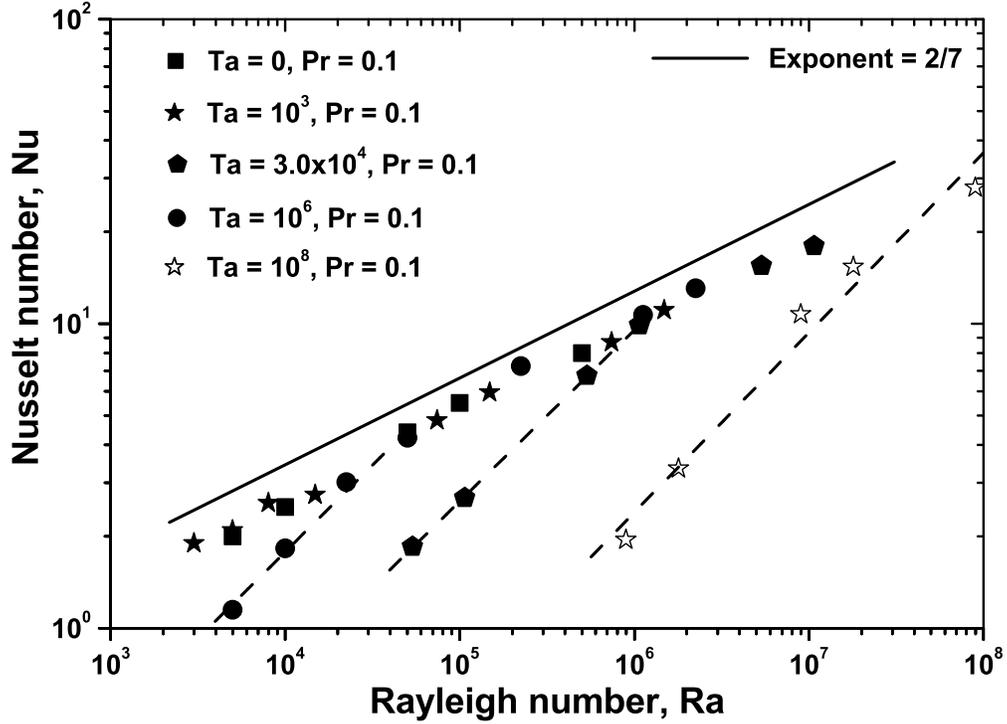}
\caption{The scaling of Nusselt number $\mathrm{Nu}$ with Rayleigh number $\mathrm{Ra}$ for diferent values of Taylor number $\mathrm{Ta}$ and Prandtl number $\mathrm{Pr}$. The scaling exponent is 2/7 for $\mathrm{Nu} > \mathrm{Nu}_c (\mathrm{Ta}, \mathrm{Pr}).$} \label{Nu_Ra} 
\end{center}
\end{figure*} 

\begin{figure*}[htbp!]
\begin{center}
\includegraphics[width=0.9\textwidth]{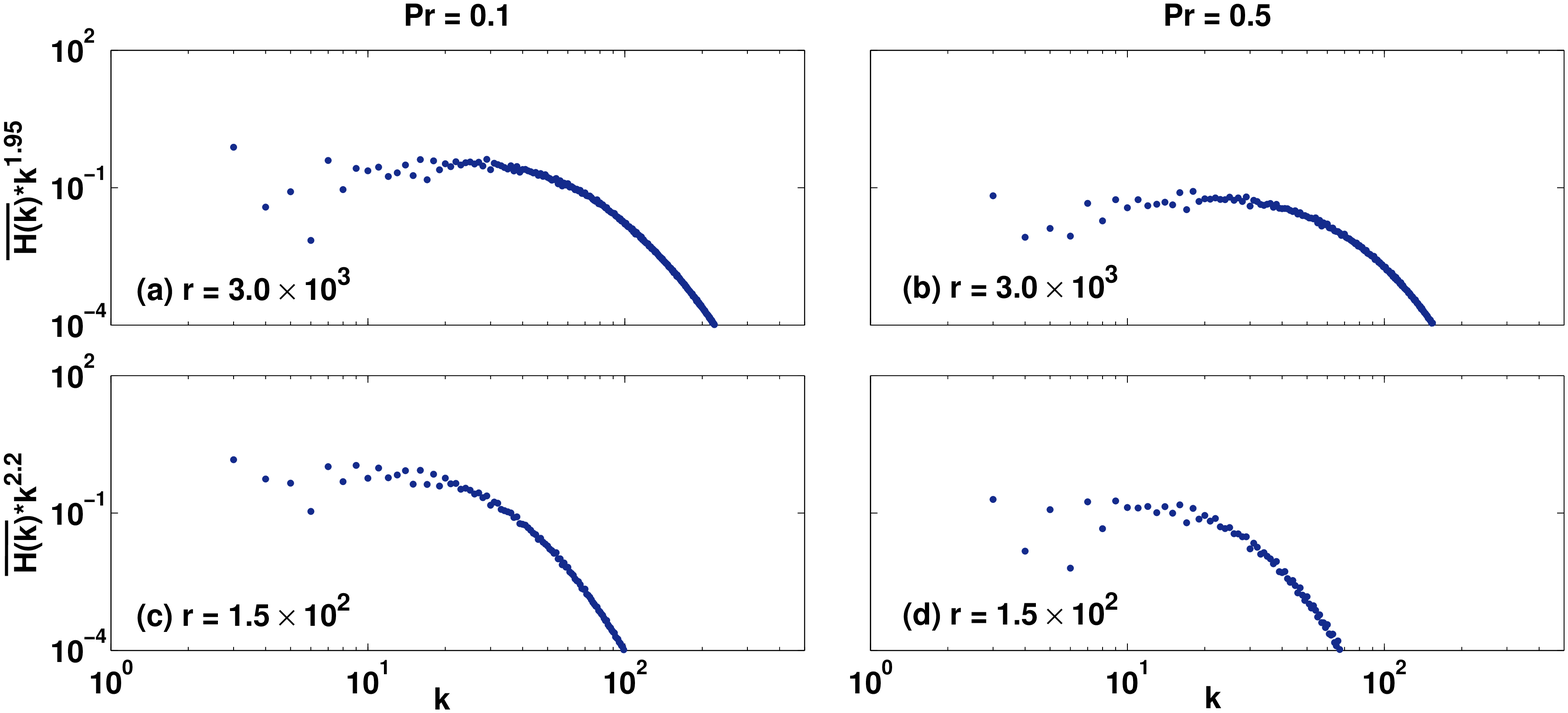}
\caption{(Color online) The compensated spectra of global convective heat flux  $\overline{\mathrm{H}(k)}*k^{\delta}$ in the absence of rotation ($\mathrm{Ta}=0$, i.e., $\mathrm{Ro}=\infty$) as computed from direct numerical simulations (DNS) for $\mathrm{Pr}=0.1$ (the left column) and for $\mathrm{Pr}=0.5$ (the right column) on $512^3$ spatial grids. The spectra are compensated by a factor $k^{1.95}$ for $r = 3.0 \times 10^3$ [(a) and (b)], while the compensating factor is $k^{2.2}$ for $r = 1.5\times 10^2$ [(c) and (d)].}\label{wt_spectra_T0} 
\end{center}
\end{figure*} 
The time averaged heat flux spectrum $\overline{\mathrm{H}(k)}$, in the absence of rotation, has been  found to scale with dimensionless wave number $k$ as $k^{-\delta}$ . The value of exponent $\delta$  has been  computed by best fitting  data points obtained from DNS. Its value has been found to be close to $1.95$ for higher value of $r$. For lower values of $r$, its value varies between $2.0$ and $2.4$.  The range of wave number and the computed values of the exponent $\delta$ are listed in Table~\ref{table1}. Figure~\ref{wt_spectra_T0} displays the compensated spectra  of the total convective heat flux $\overline{\mathrm{H}(k)}*k^{\delta}$ for low-Prandtl-number fluids ($\mathrm{Pr}=0.1$  and $\mathrm{Pr}=0.5$) in the absence of rotation ($\mathrm{Ta} = 0$). 

The time averaged spectra $\overline{\mathrm{H}(k)}$ are compensated by a factor $k^{1.95}$ [Figs.~\ref{wt_spectra_T0} (a) and (b)] for $r = 3.0 \times 10^3$. The spectra are compensated by a factor $k^{2.2}$ for  $r=1.5\times 10^2$ [Figs.~\ref{wt_spectra_T0} (c) and (d)].  The scaling behavior is more clear and the scaling range is longer for higher values of $r$. 
\begin{figure*}[hbtp!]
\begin{center}
\includegraphics[width=0.9\textwidth]{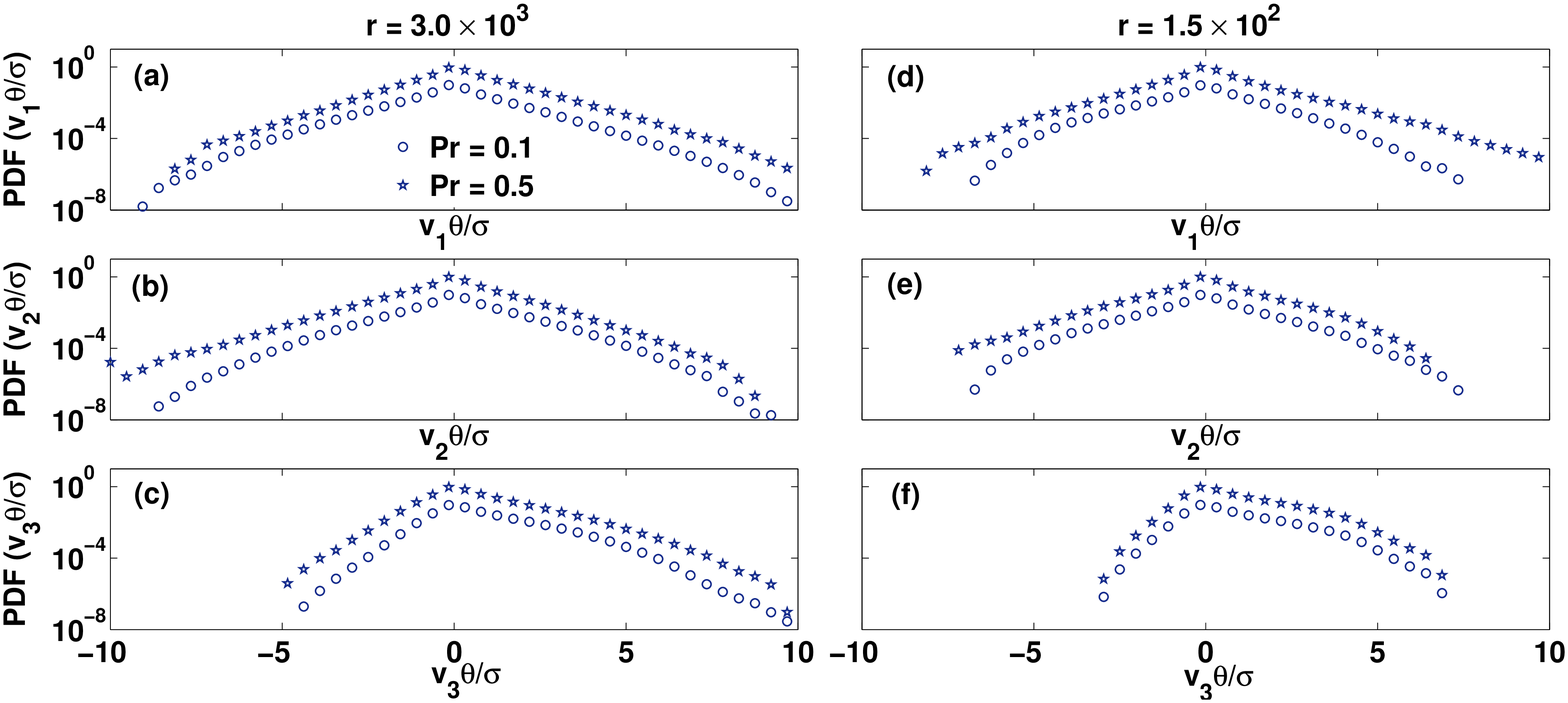}
\caption{(Color online) The time averaged value of instantaneous probability density functions (PDFs) of the convective heat fluxes in different directions in the absence of rotation ($\mathrm{Ta} = 0$, $\mathrm{Ro}=\infty$). They are computed on $256^3$ spatial grids for reduced Rayleigh number ${r} = 3.0 \times 10^3$ (the left column) and ${r} = 1.5 \times 10^2$ (the right column). The upper row  shows PDFs for $\mathrm{v}_1\theta/\sigma$ [(a) and (d)], the  middle row shows PDFs for $\mathrm{v}_2\theta/\sigma$ [(b) and (e)], and the lower row displays PDFs for $\mathrm{v}_3\theta/\sigma$ [(c) and (f)]. The curves have been shifted vertically by a factor of 10 for data points for two values of $\mathrm{Pr}$ and the heat fluxes are normalized by their standard deviations $\sigma$.} \label{pdf_thermal_fluxes_Ta0} 
\end{center}
\end{figure*}
The normalized probability distribution functions of the local heat fluxes along the horizontal and the vertical directions (for $\mathrm{Pr} = 0.1$ and $\mathrm{Pr} = 0.5$) in the absence of rotation ($\mathrm{Ta}=0$) are shown in Fig.~\ref{pdf_thermal_fluxes_Ta0}. All data points obtained from DNS on the vertical grid points for $0.2 \le z \le 0.8$   are used to compute the PDFs of instantaneous thermal fluxes in the simulation box. Several such frames are used to find the time averaged values of the PDFs of local heat fluxes in various directions. The left column shows the PDFs for $r = 3.0 \times 10^3$ and the right column shows the same for $r=1.5\times 10^2$.  The instantaneous local heat fluxes are possible in all directions and they can be either positive or negative. The probability distributions of the local heat fluxes $\mathrm{v}_1\theta/\sigma$ along the $x-$axis [Figs.~\ref{pdf_thermal_fluxes_Ta0} (a) and (d)],  $\mathrm{v}_2\theta/\sigma$ along the $y-$axis [Figs.~\ref{pdf_thermal_fluxes_Ta0} (b) and (e)],  $\mathrm{v}_3\theta/\sigma$ along vertical direction [Figs.~\ref{pdf_thermal_fluxes_Ta0} (c) and (f)] show peaks at zero. All PDFs are non-Gaussian and have exponential tails. The PDFs of the local heat fluxes $\mathrm{v}_1\theta/\sigma$ and $\mathrm{v}_2\theta/\sigma$  in the horizontal plane are symmetric about their peaks. This is due to the fact that the net heat flux in the horizontal plane is zero, although local thermal fluxes are possible in horizontal plane. All the PDFs of the  heat flux in the vertical direction are asymmetric about their peaks even in the absence of rotation ($\mathrm{Ta} = 0$) and have long tails for the  positive values of the flux.  This is consistent with the fact that a net heat flux is only in the vertically upward direction, although the thermal flux in vertically downward direction is possible locally.   This kind of behavior is also observed in the PDFs of the energy fluxes in wave turbulence~\cite{falcon_prl_2008} as well as in convective turbulence~\cite{shishkina_pof_2007,shang_prl_2003,kaczorowski_jfm_2013}. 

The scale invariance of the system of equations describing turbulence in a stratified fluid without rotation is now explored. For the most part results will agree with the existing literature as summarized in Lohse and Xia~\cite{review_RBC2}. However, this being a more general point of view where only the scale invariance of the equations of motion is imposed, it gives a complete picture of the various possibilities. The equations of motion, after elimination of the pressure fluctuations using $\partial_j v_j=0$, may be written in the dimensional form as:
\begin{eqnarray}
\partial_t{v_i} + {v_j}\partial_j{v_i}- \partial_i [{\nabla}^{-2} \partial_j  {v_l}\partial_l{v_j}] &=& \nu \nabla^{2}{v_i} + \alpha g \left(\delta_{i_{3}} - \nabla^{-2}\partial_i \partial_3\right){\delta T},\label{ns1}\\
\partial_t{\delta T} + {v_j \partial_j {\delta T}}  &=& \lambda \nabla^{2}{\delta T} + ({\Delta T}/d)v_3,\label{temp1}
\end{eqnarray}
where $v_i$ ($i =1,2,3$) stand for three components of the velocity field and $\delta T$ is the fluctuation in the temperature field due to convection. A spatial scale $l$ is introduced, which transforms distances $x_i$ to $x_i^{\prime}$ such that  $x_i=lx_i^{\prime}$. The time is assumed to scale as $t = l^{\zeta} t^{\prime}$, where $\zeta$ is the dynamical exponent expressing the slowing down the dynamics at large length scales. The velocity scales as $v_i = l^{1-\zeta} v_i^{\prime}$ as a consequence. The temperature fluctuations are assumed to scale  as $\delta T = l^{\eta} \delta T^{\prime}$. There are three system parameters $-$ $\nu, \lambda$, and $\alpha$. The kinematic viscosity $\nu$ determines the response function for the velocity field $v_i$, and the thermal diffusivity $\lambda$ determines the response function for the temperature field $\delta T$. The thermal expansion coefficient $\alpha$ is a cross-response parameter and it determines how the velocity responds to a perturbation in the temperature. In a diagrammatic perturbation theory, it is easy to see that at the lower order, the non-linearities modify the velocity and temperature response functions and not the cross-response. Hence in the Boussinesq approximation the expansion coefficient scaling may be ignored $-$ it is not true in the compressible fluid. The scale transforms lead to $\nu$ as $\nu=l^{\mu_1}\nu^{\prime}$ and $\lambda$ as $\lambda=l^{\mu_2}\lambda^{\prime}$. One demands that the dynamics in primed variables look identical to the unprimed ones if scale invariance holds and this leads to 
 \begin{eqnarray}
 \eta&=&1-2\zeta\label{eq_x}\\
\mu_1&=&\mu_2=2-\zeta\label{eq_y}
 \end{eqnarray}
No further progress is possible unless additional constraints are imposed on various fluxes. In this problem there are two kind of fluxes $-$ the energy flux $\epsilon$ and the thermal (or entropy) flux $\epsilon_S$. They are given by,
\begin{equation}
\epsilon= \frac{\partial}{\partial t}\left[\frac{1}{2V} \int v^2  d^3r\right]
\end{equation}
 and 
 \begin{equation}
 \epsilon_S= \frac{\partial}{\partial t}\left[\frac{1}{2V} \int (\delta T)^2d^3r\right]
  \end{equation}
In the absence of the temperature fluctuation, one has the usual Kolmogorov picture of turbulence in the velocity field. Kolmogorov's picture of scaling in the inertial range requires a scale independent kinetic energy flux $-$ $\epsilon$  is transformed without loss from one scale to the next. Imposition of the condition that $\epsilon$ is $k-$independent leads to $\zeta=2/3$. This gives the scaling dimension $1-\zeta$ of the velocity field as $1/3$. Knowing the scaling dimension of the $v_i$, the Kolmogorov correlations may immediately be write down as:
 \begin{eqnarray}
 [\Delta v_i (\vec{R})]^n&=&\frac{1}{V}\int d^3x [v_i(\vec{x}+\vec{R})-v_i(\vec{x})]^n
 \propto  R^{n/3}
 \end{eqnarray}
What about the $5/3 $ law of Kolmogorov? That pertains to $E(k)$, which is obtained from  the wave number dependent velocity $v_i(k)$ through the correlation function $C(k)=\langle v_i(k)v_i(-k)\rangle$ when one looks at the total energy per unit mass $E$ which can be written as
\begin{equation}
 E=\int \frac{d^3k}{(2\pi)^3} \langle v_i(k)v_i(-k)\rangle=\int \frac{d^3k}{(2\pi)^3} C(k)=\int dk E(k)
 \end{equation} 
so that, 
\begin{equation}
E(k)= \int \frac{4\pi k^2}{(2\pi)^3} C(k).\label{eq_ck}
\end{equation}
$C(k)$ may be found from a scaling argument since it is the Fourier transform of the translationally invariant spatial correlation \cite{kumar_2014}
$\langle v_i(\vec{x})v_i(\vec{x}+\vec{R})\rangle$, which from the discussion of scaling dimension has to be proportional to $R^{2/3}$. Using the Fourier transform $C(k)\propto k^{-(3+2/3)}$  in Eq.(\ref{eq_ck}), one arrives at Kolmogorov scaling law~\cite{K41}: $E(k)\propto k^{-5/3}$. The scale invariance argument gives the wave number dependence directly without giving the dependence on $\epsilon$. What does the scaling invariance imply? It means that there exists a re-normalization group fixed point (in the absence of anomaly $-$ in this case intermittency) which corresponds to the behavior of $E(k)$. This is precisely the content of Yakhot and Orszag~\cite{yakhot}, where apparently the scale invariance of $\epsilon$ is not involved. It should be noted that these authors do not have to invoke it, as is done in this article, because they make an alternative assumption that the random forcing $f$ is prescribed such that the second moment $\langle f(k)f(-k)\rangle$ falls off as $k^{-D}$ in a $D$-dimensional space. This assumption was necessary for any analytic treatment of correlation.

Having clarified the meaning and strength of invoking scale invariance, we now note that in the case of the stratified fluid, there may have three different possibilities of scale invariant fluxes:\\
(I) $\epsilon$ is $k$-independent but $\epsilon_S$ is not $-$ this is called here as K41 for the very specific reason that in K41 there is only one scale independent transfer and that is of the kinetic energy.\\
(II) $\epsilon_S$ is $k$-independent but $\epsilon$ is not.  This is Bolgiano-Obukhov scenario, which is denoted here as BO.\\
(III) $\epsilon$ and $\epsilon_S$ are both scale independent, a situation which is termed as KBO in this article. This is conventionally called K41 in the literature and what is called as K41 here has never been noticed, as the general scale invariance on equations of motion was never imposed. Each of these cases is discussed separately below.\\\\
(I) The constancy of $\epsilon$ leads to $\zeta=2/3$ and immediately $\eta=-1/3$ from Eq. (\ref{eq_x}). Looking for the $k-$dependence of $\epsilon_S$, one immediately finds that $\epsilon_S \propto k^{4/3}$, i.e., $\epsilon_S$ rises as $k$ increases. The entropy spectrum $S (k)$ defined as
\begin{equation}
\int S(k)dk=\frac{1}{V}\int \langle \delta T(k) \delta T(-k)\rangle \frac{d^3k}{(2\pi)^3}
\end{equation}
is consequently seen to behave as (based on $\eta=-1/3$) $S                                                                                                                                                                                                                                                                                                                                                                                                                                                                                                                                                                                                                                                                                                                                                                                                                                                                                                                                                                                                                        (k)\propto k^{-1/3}$.\\\\
(II) In this range, the $k$-independence of $\epsilon_S$ leads to $\zeta=2\eta$ and from Eq. (\ref{eq_x}), it yields $\eta=1/5$ and $\zeta=2/5$. It follows that $\epsilon \propto k^{-4/5}$, i.e., $\epsilon$ decreases as $k$ increases. The scaling dimension of $v_i(\vec{R})$ is $1-\zeta=3/5$ and it follows that 
\begin{equation}
E(k)\propto k^{-11/5}\label{eq_BO1}\\
\end{equation}
while 
\begin{equation}
S(k) \propto k^{-7/5}\label{eq_BO2}
\end{equation}
(III) In this range, $\epsilon$ and $\epsilon_S$ are both $k-$independent. This leads to $\zeta=2\eta$ and $\zeta=2/3$ simultaneously, which gives $\eta=1/3$. The velocity fields and temperature have the same scaling dimension of $1/3$ and hence 
\begin{equation}
E(k)\propto k^{-5/3}\\
\end{equation}
and 
\begin{equation}
S(k) \propto k^{-5/3}
\end{equation}

It should be pointed out that if a real perturbative re-normalization calculation is to be carried out to obtain the scaling field points, then it is a meaningful perturbation theory only for $\Delta T<0$, i.e., for the stable stratification. While the above scale invariance arguments are carried out with no heed to the sign of $\Delta T$, a real calculation backup is feasible only for $\Delta T<0$. What have extensive numerical analysis on stratified flows revealed? The most clear cut answers are in Ref.~25, i.e., in the BO regime for $\Delta T<0$. This is where it has been clearly demonstrated that there exists a regime where $\epsilon_S$ is constant and $\epsilon \propto k^{-4/5}$. The spectra are as given in Eqs.~(\ref{eq_BO1}) and (\ref{eq_BO2}). It has also been shown recently~\cite{jkb_2015} that this regime is obtained for low values of Richardson number $\mathrm{Ri}=[\alpha (\Delta T) gd]/{\overline{v^2}}$. As the Richardson number is varied in this work there is, for $\mathrm{Ri}<<1$, a flat $\epsilon$ with $E(k)\propto k^{-5/3}$. Unfortunately, no comment on $\epsilon_S$ and $S(k)$ are actually available in this range.

Apart from the Ref.~25, a systematic study of the fluxes as a function of the wave number are never reported before. As this picture is not investigated earlier, it is impossible to ascertain correctly the status of the much more extensive studies on $E(k)$ and $S(k)$. It is worth to point out that all these spectra will necessarily exhibit crossover behavior as a given situation cannot separate the three regions discussed. They have to exist simultaneously and one has to look for wave numbers where one contribution dominates the other. This is only natural as no one switches off one flux or the other in studying the various domains. It is a question of dominance. It has always been assumed that there is one scale $k_B -$ the Bolgiano scale which dictated where the crossover should occur. Clearly if $k<<k_B$, it is BO scaling and $k>> k_B$ it is KBO/K41 (arguments have always been for $E(k)$ crossover and not for $S(k)$ crossover), where the dimensionless  Bolgiano wave number $k_{B} = 2\pi d/L_B$ corresponding to the  global Bolgiano length $L_B/d = \left(\mathrm{Nu} \right )^{1/2}/{(\mathrm{Ra} \mathrm{Pr})}^{1/4}$. Using a dimensional analysis, one arrives at $E(k) = C_E\epsilon^{2/3}k^{-5/3}$ and $E(k)=C_S \epsilon_S^{2/5}(\alpha g)^{4/5}k^{-11/5}$, where $C_E$ and $C_S$ are numbers of $O(1)$. The scale $k_B$ comes from
\begin{equation}
C_S \epsilon_S^{2/5}(\alpha g)^{4/5}k_B^{-11/5}=C_E \epsilon^{2/3}k_B^{-5/3}
\end{equation}
or
\begin{equation}
k_B^{8/15}=\frac{C_S}{C_E}\frac{\epsilon_S^{2/5}}{\epsilon^{2/3}}(\alpha g)^{4/5}
\end{equation}

The numbers $C_E$ and $C_S$ can actually be calculated and some of them exist in literature. In the Kolmogorov scenario, the relaxation rate is written as $\Gamma_1 k^{2/3}$ and the correlation function as $C_1 k^{-11/3}$. A self-consistent one loop evaluation of the relaxation rate gives $\Gamma_1/C_1 =I_E$ where $I_E$ is the numerical value of the one loop integral. The energy transfer $\epsilon$ is the three point correlation function which has been carefully evaluated in Leslie~\cite{leslie:book_1973} and $\epsilon =I C_1^2/\Gamma_1$ where $I$ is the numerically evaluated integral. Combining with one loop relaxation rate, one arrives at $C_1= \left( I_E^{1/3}/I^{2/3}\right)\epsilon^{2/3}$, which identifies $C_E$ as the number $I_E^{1/3}/I^{2/3}$. Similar calculation may be done in the Bolgiano regime, where the relaxation rate is $\Gamma_2 k^{2/5}$ and the correlation function as $C_2 k^{-21/5}$. One now needs to evaluate transfer rate $\epsilon_S$ of the convective entropy from the three point correlation function obtained from Eq.~\ref{temp1}. The additional feature of the entropy correlation function is now required, which is written as $C_2^{\prime} k^{-17/5}$ in the Bolgiano regime and the entropy relaxation rate is taken as $\Gamma_2^{\prime} k^{2/5}$. One evaluates the three point function for $\epsilon_S$ as $\epsilon_S = J_{2/5} C_2 C_2^{\prime}/\Gamma_2$, where $J_{2/5}$ is a numerically computed result of an integral. The correlation coefficients $C_2$ and $C_2^{\prime}$ are related through the linear terms in Eq.~\ref{ns1} by a factor $f$ and one needs the one loop integral $I_B$ from Eq.~\ref{temp1} to find $C_2=\frac{\epsilon_S^{2/5}}{f^{2/3}} I_B^{-1/5} (J_{2/5})^{-2/5}$ which identifies $C_S$ as the number $(I_B f J_{2/5})^{-2/5}$. The Bolgiano wave number above refers to the kinetic energy spectrum. What about the entropy spectrum? In KBO regime the entropy spectrum is $C_1^{\prime}k^{-11/3}$, while in the BO regime it is $C_2^{\prime}k^{-17/5}$. The crossover wave number will be determined when  two contributions are equal. The coefficients $C_1^{\prime}$ and $C_2^{\prime}$ are expressed in terms of a different set of numbers $-$ in particular $J_{2/5}$ will be changed to $J_{2/3}$ as the relaxation rate for calculating the integral in the entropy transfer integral and the factor $f$ will be absent. The cross-over wave number estimated here is a global one. It is different from the results of Calzavarini et al.~\cite{calzavarini_etal_2002} who estimated cross-over lengths at different fluid heights. 

It is clear that a difference of factor $3$ between $C_S$ and $C_E$ leads to a difference of one order magnitude in $k_B$ from the naive assumption of $C_S$ and $C_E$. The lesson is that the Bolgiano wave number $k_B$ calculated from $E(k)$ can differ easily by an order of magnitude from $k_B$ calculated from $S(k)$, since the numerical constants associated are different from what one has for $E(k)$. Thus the transition from $k^{-5/3}$ to $k^{-11/5}$ for $E(k)$ can happen in a very different $k_B$ from the transition from $k^{-5/3}$ to $k^{-7/5}$ for $S(k)$. 

Similarly for the region I, where one finds $S(k) \sim k^{-1/3} $, it is very likely that it can never be seen in an entropy spectrum as for small wave numbers it will be swamped by both $k^{-7/5}$ and $k^{-5/3}$, but it is possible that the existence of this region may be indicated by an increase of $\epsilon_S$ with increasing $k$ if the thermal flux is carefully measured.

With the above argument, one may now immediately write down the spectrum $\mathrm{H}(k)$ for the correlation function $\langle v_i(k) \delta T(-k)\rangle$ as 
\begin{eqnarray}
&Region&~(I)~~ \mathrm{H}(k) \propto  k^{-1}\\
&Region&~(II)~~ \mathrm{H}(k) \propto k^{-9/5}\\
&Region&~(III)~~ \mathrm{H}(k) \propto k^{-5/3}
\end{eqnarray}
It should be noted that according to the naive scaling discussed so far $\mathrm{H}(k)\propto [E(k)S(k)]^{1/2}$. The results obtained by various authors on $E(k)$ and $S(k)$ are explored now. As for $E(k)$, it is always in some crossover range between $k^{-5/3}$ and $k^{-11/5}$ but $S(k)$ data is more specific. It shows two clear cut features:\\
(i) there is always a clear $k^{-7/5}$ and no trace of $k^{-5/3}$ and\\
(ii) for $k$-values less than some critical value, there is a bifurcation and the $S(k)$ vs $k$ curve has two branches in the absence of rotation~\cite{mishra_etal_pre_2010} $-$ one scaling as $k^{-7/5}$ and the other as $k^{-2}$. The spectrum of convective entropy $S(k)$ in the presence of rotation also shows two branches\cite{hkp_kk_jkb_pre1_2014}. 
One finds the scaling for one branch as $k^{-7/5}$ and for the other branch the exponent  varies from $2.8$ to $3.8$. In Ref.~14 there is an argument about why the spectrum should be $k^{-2}$ for a specific set of modes. By analyzing their numerical data on mode to mode entropy transfer, they conclude that the status of $\delta T(0,0,2n)$ modes are different and it is for these modes that $S(k)$ is $k^{-2}$. The numerics in Ref.~29 show the same bifurcation as in Ref.~14 but with a steeper slope than $k^{-2}$, which could be a consequence of finding the exponent by the best fit. However,  the temperature modes that contribute to the upper branch of the entropy spectrum are of the form $0,0,2n$. The coupling of these temperature modes to the vertical velocity modes does not yield a nonzero value of $<v_3 \delta T>$ after averaging over the whole simulation box. Hence, a pure Bolgiano scaling would always give $k^{-9/5}$ for the global heat flux. The slightly larger value of the measured exponent is presumably due to a cross-over effects. The noise is always more at lower values of $k$ as the number of points are less in  smaller spherical shells.

The probability distribution functions of the local thermal fluxes in the simulation box are now discussed. Following Falcon et al.~\cite{falcon_prl_2008}, one can model a system where the fluctuations in the vertical velocity $v_3$ are driven by fluctuations in the convective temperature field $\theta$, i.e., $\dot{v_3} + \Gamma v_3 = \theta $, where $\Gamma$ stands for viscous dissipation. The thermal fluctuations decay due to thermal diffusion but are maintained in the model by random force $\xi$ which is a Gaussian white noise, i.e., $\dot{\theta} + \gamma \theta = \xi(t)$, where $<\xi (t_1)\xi (t_2)>= D\delta(t_1-t_2)$. The PDF of $v_3$ and $\theta$ is a bi-variate normal distribution. To get the distribution of $v_3\theta$, one writes $F=v_3\theta$ and integrates over $\theta$ as by Falcon et al.~\cite{falcon_prl_2008} to obtain

\begin{figure}[h!]
\begin{center}
\includegraphics[width=\textwidth]{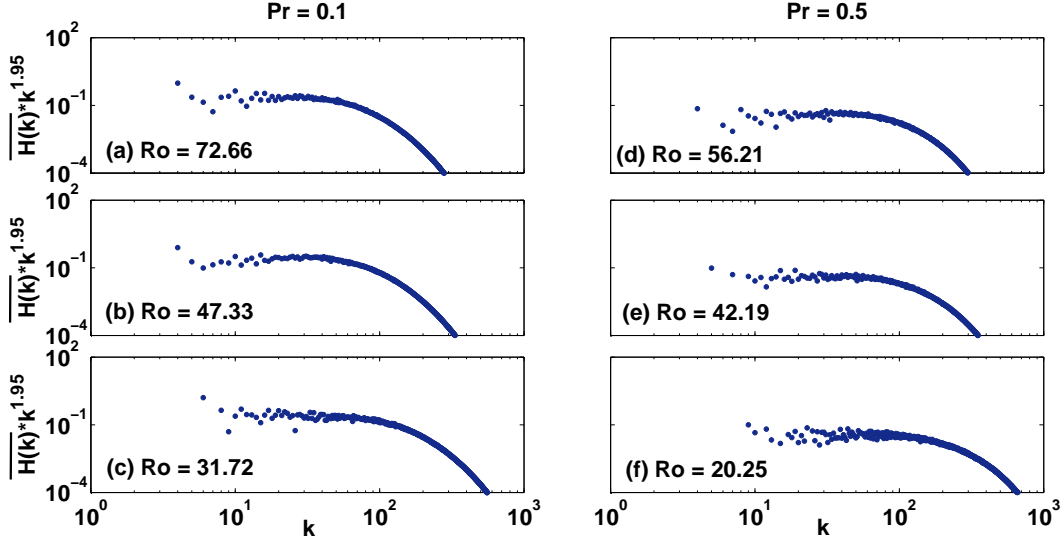}
\caption{(Color online) Compensated spectra $\overline{\mathrm{H}(k)}*k^{1.95}$ of the global convective heat flux for larger values of Rossby number $\mathrm{Ro}$. They are computed on $512^3$ spatial grids for $\mathrm{Pr}=0.1$ (the left column) and $\mathrm{Pr}=0.5$ (the right column) for (a) $\mathrm{Ro} = 72.66$, (b) $\mathrm{Ro} = 47.33$, (c) $\mathrm{Ro} = 31.72$, (d) $\mathrm{Ro} = 56.21$, (e) $\mathrm{Ro} = 42.19$, and (f) $\mathrm{Ro} = 20.25$. The corresponding  values of $\mathrm{Ta}$, $\mathrm{Pr}$ and $r$ corresponding to different values of $\mathrm{Ro}$ are given in Table~\ref{table1}}. 
\label{wt_spectra_r3000} 
\end{center}
\end{figure} 
\begin{figure}[t!]
\begin{center}
\includegraphics[width=\textwidth]{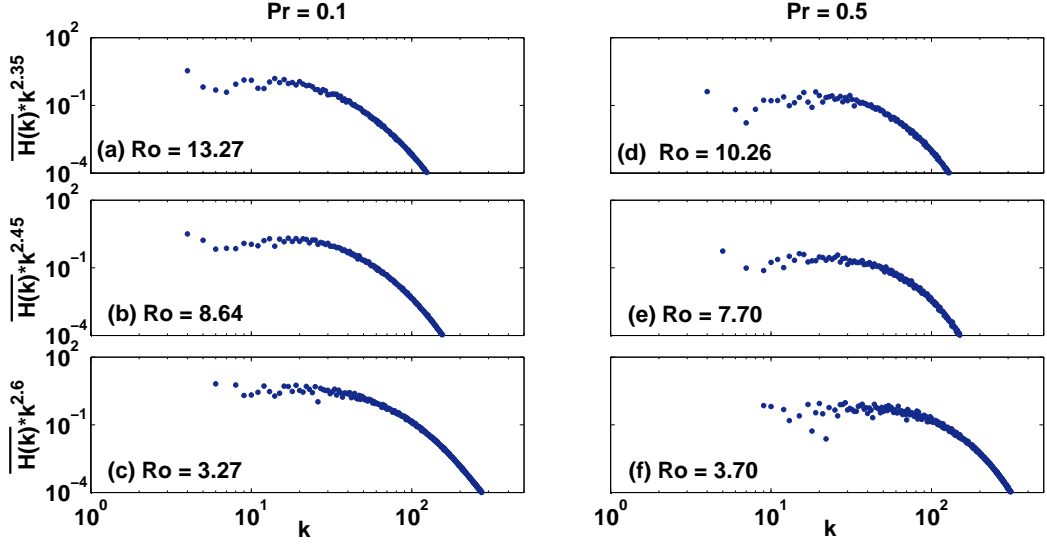}
\caption{(Color online) Compensated spectra of the global heat flux $\overline{\mathrm{H}(k)}*k^{\delta}$ in different directions for relatively smaller values of $\mathrm{Ro}$. They are computed on $512^3$ spatial grids for $\mathrm{Pr}=0.1$ (the left column) and $\mathrm{Pr}=0.5$ (the right column). The possible values of the exponent $\delta$ is computed from the best fit of DNS data. The value of $\delta$ is taken equal to $2.35$ for the upper row [(a) $\mathrm{Ro} = 13.27$ and (d) $\mathrm{Ro} = 10.26$],  $2.45$ for the middle row [(b) $\mathrm{Ro} = 8.64$ and (e) $\mathrm{Ro} = 7.70$] and  $2.6$ for the lower row [(c) $\mathrm{Ro} = 3.27$ and (f) $\mathrm{Ro} = 3.70$]. The values of $\mathrm{Ta}$, $\mathrm{Pr}$ and $r$ corresponding to different values of $\mathrm{Ro}$ used here are listed in Table~\ref{table1}.} \label{wt_spectra_r100} 
\end{center}
\end{figure}
\begin{figure}[t!]
\begin{center}
\includegraphics[width=0.9\textwidth]{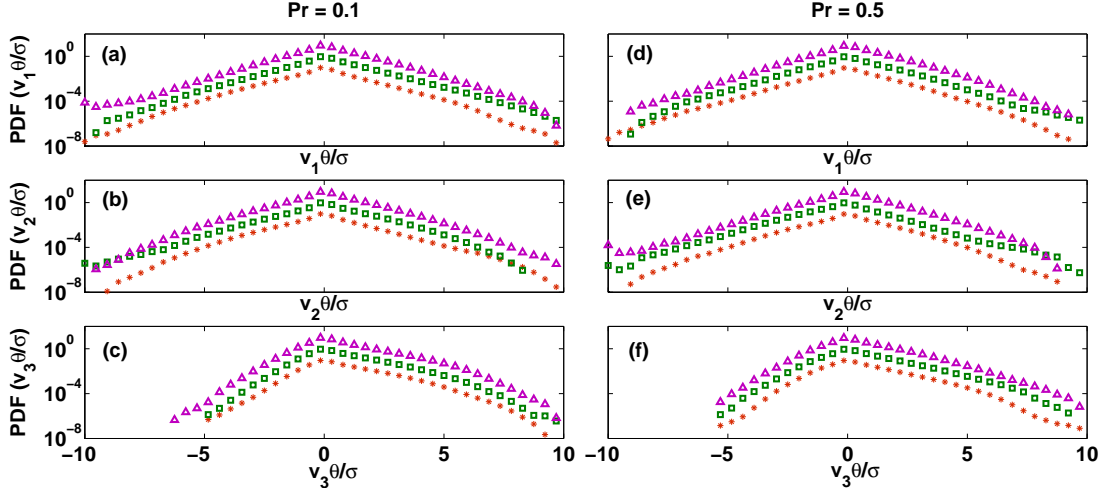}
\caption{(Color online) The time averaged value of the instantaneous probability density functions (PDFs) of the local convective heat fluxes for $\mathrm{Pr} = 0.1$ (the left column) and for $\mathrm{Pr} = 0.5$ (the right column) computed on $256^3$ spatial grids and $r = 3.0 \times 10^3$. PDFs for $\mathrm{v}_1\theta/\sigma$ are shown in the upper row [(a) and (d)], for $\mathrm{v}_2\theta/\sigma$ in the middle row [(b) and (e)] , and for $\mathrm{v}_3\theta/\sigma$ in the lower row [(c) and (f)]. Data points for 
$\mathrm{Ta} = 10^4$, $\mathrm{Ta} = 3.0 \times 10^4$, $\mathrm{Ta} = 10^6$ are shown by red asterisks, green rectangles, and magenta triangles, respectively. The curves are shifted vertically by a factor of $10$ for each value of $\mathrm{Ta}$ for clarity.} \label{pdf_thermal_fluxes_Pr01_r3000} 
\end{center}
\end{figure}

\begin{figure}[t!]
\begin{center}
\includegraphics[width=0.9\textwidth]{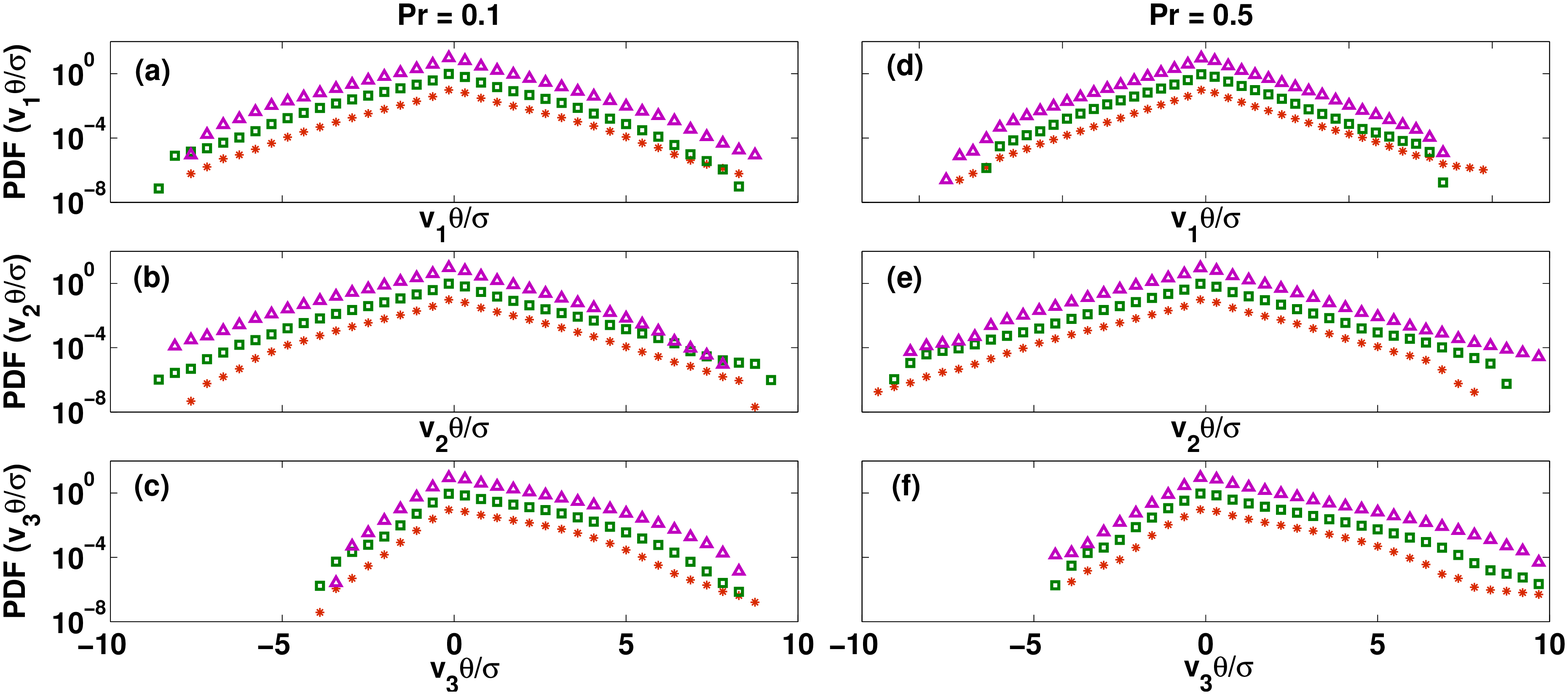}
\caption{(Color online) The time averaged value of the instantaneous probability density functions (PDFs) of the local convective heat fluxes for $\mathrm{Pr} = 0.1$ (the left column) and for $\mathrm{Pr} = 0.5$ (the right column) computed on $256^3$ spatial grids for $r=10^2$. PDFs for $\mathrm{v}_1\theta/\sigma$, $\mathrm{v}_2\theta/\sigma$ and $\mathrm{v}_3\theta/\sigma$ are displayed in the upper, the middle and the lower rows, respectively. Data points for $\mathrm{Ta} = 10^4$, $\mathrm{Ta} = 3.0 \times 10^4$, $\mathrm{Ta} = 10^6$ are shown by red asterisks, green rectangles, and magenta triangles, respectively. The curves are shifted vertically by a factor of $10$ for each value of $\mathrm{Ta}$ for clarity.} \label{pdf_thermal_fluxes_Pr01_r100} 
\end{center}
\end{figure}

\begin{equation}
P(F)=\frac{\exp\left[\frac{\tau F}{(1-\tau^2)\sigma_1 \sigma_2}\right]}{\pi\sigma_1\sigma_2\sqrt{1-\tau^2}}K_0\left[ \frac{F}{(1-\tau^2)\sigma_1\sigma_2} \right] \label{eqn_a}
\end{equation}
 
where $\sigma_1^2 = \frac{D}{2\Gamma \gamma(\Gamma +\gamma)}, \sigma_2^2 = \frac{D}{2\gamma}, \tau^2=\frac{\Gamma}{\Gamma +\gamma}$ and $K_0$ in the zeroth order modified Bessel function of the second kind. Using the asymptotic form of $K_0(X)$, one arrives at  $P(X)\sim \frac{1}{\sqrt{X}}e^{\tau X}e^{-|X|}$, where $X=\frac{F}{(1-\tau^2)\sigma_1\sigma_2}$. This is clearly asymmetric about $X=0$ with a sharp fall off for $X<0$.


\section{Effects of Rotation}
 The convective flow with the system rotating about the $z$-axis with a uniform speed $\Omega$ is now considered. The equation of motion for the velocity field becomes
\begin{equation}
\partial_t{v_i} + {v_j}\partial_j{v_i}- \partial_i [{\nabla}^{-2} \partial_k {v_l}\partial_l{v_k}] = \nu \nabla^{2}{v_i}  
- 2 \Omega(\epsilon_{i3k} v_k + \nabla^{-2}\partial_i \omega_3)
+\alpha g \left(\delta_{i_{3}} - \nabla^{-2}\partial_i \partial_3\right){\delta T},\label{ns2}
\end{equation}
where $\omega_3=\partial_x v_2 - \partial_y v_1$ is the vertical vorticity.  The uniform rotation about a vertical axis couples the vertical velocity to the vertical vorticity. It is the coupling to the vorticity field through the angular velocity and the fact that the externally imposed $\Omega$ should not transform under scale transformation which render useless the arguments of Sec. II to obtain scaling laws. One may only argue that for small rotation speeds (high Rossby numbers) there will be corrections to scaling and the deviation from the scaling exponents of Sec. II will increase with increasing rotation speed at fixed Rayleigh number $\mathrm{Ra}$ below a critical value. The results of simulations with rotation are presented below.

The compensated spectra of the total convective heat flux $\overline{\mathrm{H}(k)}*k^{\delta}$ for low-Prandtl-number fluids in the presence of rotation is shown in Fig.~\ref{wt_spectra_r3000}. The spectra shown in the left column are for $\mathrm{Pr}=0.1$ [Figs.~\ref{wt_spectra_r3000} (a)-(c)] and those displayed in the right column are for $\mathrm{Pr} =0.5$ [Figs.~\ref{wt_spectra_r3000} (d)-(f)] at higher values of $r$ ($\geq 3.0\times 10^3$) for relatively higher values of Rossby number ($\mathrm{Ro}>20$). The corresponding values of $\mathrm{Ta}$, $\mathrm{Pr}$ and $r$ are listed in  Table~\ref{table1}. The heat flux spectra $\overline{\mathrm{H}(k)}$ have been found to scale with dimensionless wave number $k$ as $k^{-\delta}$. The scaling exponent $\delta$ is found to be approximately equal to $2.0$. The exponent is almost independent of $\mathrm{Ta}$ and $\mathrm{Pr}$ in the range of $\mathrm{Ra}$ where the universal scaling~\cite{pharasi_etal_pre_2011} $\mathrm{Nu}\sim\mathrm{Ra}^{2/7}$ holds. The scaling behavior is more clear and the scaling range is longer in the presence of rotation. The scaling regime is found to shift towards higher values of $k$ with an increase in $\mathrm{Ta}$. The range of wave numbers and computed values of the scaling exponent $\delta$ are given in Table~\ref{table1}.

Figure~\ref{wt_spectra_r100} displays the compensated spectra of the total convective heat flux $\overline{\mathrm{H}(k)}$ for  $r=10^2$ in lower-Prandtl-number fluids ($\mathrm{Pr}=0.1$ and $0.5$) for different values of Taylor number $\mathrm{Ta}$. This is the case for relatively low values of the Rossby number ($1 < \mathrm{Ro} < 14$). The left column shows the spectra for  $\mathrm{Pr}=0.1$ [Figs.~\ref{wt_spectra_r100} (a)-(c)] and right for $\mathrm{Pr} = 0.5$ [Figs.~\ref{wt_spectra_r100} (d)-(f)] for $r = 10^2$. The corresponding values of  $\mathrm{Ta}$, $\mathrm{Pr}$ and $r$ are listed in Table~\ref{table1}. In this range, the scaling exponent $\delta$ obtained from the best fit is found to vary from $2.2$ to $2.7$. The exponent $\delta$ in this case depends upon  $\mathrm{Ta}$ and  $\mathrm{Pr}$. The value of exponent $\delta$ increases slightly with an increase in $\mathrm{Ta}$. The range of wave numbers and the best fit values of $\delta$ are listed in Table~\ref{table1}.

Normalized PDFs of the thermal flux in the presence of rotation are shown in Fig.~\ref{pdf_thermal_fluxes_Pr01_r3000}. They display the probability distributions of the thermal fluxes in the horizontal and the vertical planes for $\mathrm{Pr} = 0.1$ [Figs.~\ref{pdf_thermal_fluxes_Pr01_r3000} (a)-(c)] and $\mathrm{Pr} = 0.5$ [Figs.~\ref{pdf_thermal_fluxes_Pr01_r3000} (d)-(f)] at higher values of $r$ ($\geq 3.0 \times 10^3$) for different values of $\mathrm{Ta}$. Similar to non-rotating case, PDFs of the local heat fluxes are computed in the central part ($0.2 \le z \le 0.8$) of the simulation cell, which is away from the boundaries. The probability distributions of $\mathrm{v}_1\theta/\sigma$, $\mathrm{v}_2\theta/\sigma$, and $\mathrm{v}_3\theta/\sigma$ also show peaks at zero. PDFs are symmetric for horizontal heat fluxes but asymmetric for the vertical flux, same as seen for non-rotating case. The asymmetric shapes of the PDFs for the vertical flux have exponential tails. The area under the PDFs for  $\mathrm{v}_1\theta/\sigma$ and $\mathrm{v}_2\theta/\sigma$ in the simulation cell is zero even for finite values of $\mathrm{Ta}$.

Figure~\ref{pdf_thermal_fluxes_Pr01_r100} shows the PDFs of the  heat fluxes at $r = 10^2$ for different values of $\mathrm{Ta}$. The basic features of PDFs for lower values of $r$ remain similar to those observed at higher values of $r$. The area under the PDFs for heat flux in the horizontal plane remains zero, while the same for the  heat flux in the vertical direction is always finite and positive.  All  PDFs for local heat fluxes at lower values of the reduced Rayleigh number also show roughly  exponential tails.

\section{Conclusions}
The spectrum of heat flux in wave number space has been investigated numerically. The spectrum of the heat flux $\mathrm{H}(k)$ scales with wave number $k$ as  $k^{-{\delta}}$ with $\delta \approx 2$ for larger values of the Rossby number ($\mathrm{Ro} > 20$). For smaller values of the Rossby number ($1 < \mathrm{Ro} < 15$) the value of the scaling exponent $\delta$ increases and its value is found to vary between $2.2$ and  $2.6$.	PDFs of the thermal fluxes in different directions in the central region of the cell are non-Gaussian with their peaks at zero. The PDF of the thermal flux in horizontal direction is symmetric about its peak showing zero net flux in the horizontal direction. The PDF of the local flux in the vertical direction is asymmetric about its peak showing a net flux in vertically upward direction. All PDFs with or without rotation have roughly two exponential tails. This kind of behavior was also observed by Shishkina and Wagner~\cite{shishkina_pof_2007} in local fluxes in cylindrical geometry in the absence of rotation.

\noindent{\bf ACKNOWLEDGEMENTS}\\
We have benefited greatly with fruitful discussions with Stephan Fauve and Priyanka Maity. 

\end{document}